\documentstyle[twocolumn]{jpsj2}
\def\vS{{\mib S}}
\def\vT{{\mib T}}

\def\runauthor{Akiyuki {\sc Tokuno},Kiyomi {\sc Okamoto}}
\title{Inversion phenomenon and phase diagram of the $S=1/2$ distorted
	diamond chain with the $XXZ$ interaction anisotropy} 
\author{Akiyuki \textsc{Tokuno}\thanks{E-mail :
toku@stat.phys.titech.ac.jp} and Kiyomi \textsc{Okamoto}\thanks{E-mail : kokamoto@stat.phys.titech.ac.jp}}
\inst{Department of Physics, Tokyo Institute of Technology, Meguro-ku, Tokyo 152-8551}
\recdate{November 6, 2000}
\abst{
 We discuss the anisotropies of the Hamiltonian and the
 wave-function in an $S=1/2$ distorted diamond chain.  The
 ground-state phase diagram of this model is
 investigated using the degenerate perturbation theory up to the
 first order and the level spectroscopy analysis of the numerical
 diagonalization data.  In some regions of the obtained phase
 diagram, the anisotropy of the Hamiltonian and that of the
 wave-function are inverted, which we call {\it inversion
 phenomenon}; the N\'{e}el phase appears for the $XY$-like
 anisotropy and the spin-fluid phase appears for the Ising-like
 anisotropy.  Three key words are important for this nature,
 which are frustration, the trimer nature, and the $XXZ$
 anisotropy.}
	 
 \kword{spin chain, quantum phase transition, degenerate
 perturbation theory, frustration, trimer}

\begin{document}
 \maketitle
 
 \section{Introduction}
\begin{figure}
  \begin{center}
   \scalebox{0.6}[0.6]{\includegraphics{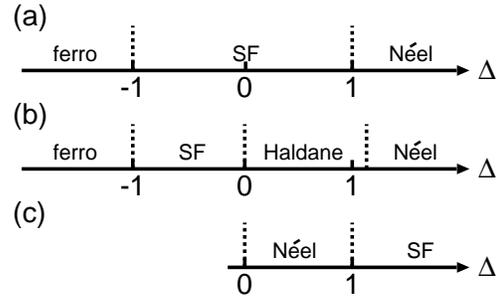}}
   \caption{Schematic diagrams of the relation between the ground state
   and anisotropy for (a) a simple $S=1/2$ $XXZ$ chain,
   (b) a simple $S=1$ $XXZ$ chain, and (c) a DD chain for certain parameter
   sets.  Here, $\Delta$ denotes the anisotropy (see Eq.(\ref{XXZ})), and ferro, SF, and Haldane
   denote the ferromagnetic state, the spin-fluid state, and Haldane
   state, respectively.}
   \label{pd_spinchain}
  \end{center}
 \end{figure}

 In quantum spin chains, a value of the anisotropy of the $XXZ$ type
 interaction is an important factor regarding to what kind of the
 ground state is realized.  In a case of an $S=1/2$ $XXZ$ chain, for
 instance, the Ising-like anisotropy brings about the N\'{e}el or
 ferromagnetic ground state, and the $XY$-like anisotropy brings about the spin-fluid (SF)
 ground state (see Fig.\ref{pd_spinchain}(a)).  In a case of an $S=1$
 $XXZ$ chain, the Ising-like anisotropy also brings about the N\'{e}el or
 ferromagnetic ground state (see Fig.\ref{pd_spinchain}(b)).  However, this relation between the anisotropy
 and the ground state is sometimes broken for frustrated systems.
 In fact, a novel nature in the ground-state phase diagram has been
 found by Okamoto and Ichikawa\cite{Okamoto1}.   That is, 
 this relation is inverted in an $S=1/2$ distorted diamond (DD) chain\cite{DD1,DD2,DD3}
 with the $XXZ$ type interaction.  Namely, the SF state appears for the Ising-like
 anisotropy in some regions of the phase diagram of
 the ground state, and the N\'{e}el state appears for the $XY$-like
 anisotropy (see Fig.\ref{pd_spinchain}(c)).  This inversion between the anisotropy of the Hamiltonian
 and that of the wave-function in the ground state is called the {\it inversion
 phenomenon}.  From the viewpoint of the quantum statistical physics,
 this phenomenon is novel and exotic.  It is found by one of the present authors (K.O.) that the inversion phenomenon also
 appears in an $S=1/2$ trimerized $XXZ$ chain with the
 next-nearest-neighbor interaction.\cite{Okamoto2} 
 
\begin{figure}
  \begin{center}
   \scalebox{0.4}[0.4]{\includegraphics{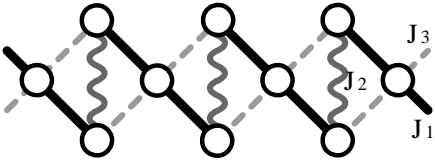}}
   \caption{A sketch of the DD chain: solid lines, wavy lines, and dotted lines
   denote $J_1$, $J_2$, and $J_3$ coupling bond, respectively.}
   \label{fig:1}
  \end{center}
 \end{figure}

 In this paper, we elucidate the origin of the inversion
 phenomenon.  We discuss the generalized DD chain model
 with the three kinds of the anisotropy parameters sketched in
 Fig.\ref{fig:1}.   The Hamiltonian is
 \begin{eqnarray}
  {\cal H} &=& J_{1}\sum_{j}\{
   \vS_{3j-1}\cdot\vS_{3j}(\Delta_{1})+\vS_{3j}\cdot\vS_{3j+1}(\Delta_{1})\}
   \nonumber \\
  & & +J_{2}\sum_{j}\vS_{3j+1}\cdot\vS_{3j+2}(\Delta_{2}) \nonumber \\
  & & +J_{3}\sum_{j}\vS_{3j}\cdot\vS_{3j+2}(\Delta_{3}) \label{H},
 \end{eqnarray}
 where
 \begin{equation}
  \vS_{i}\cdot\vS_{j}(\Delta) \equiv
   S_{i}^{x}S_{j}^{x}+S_{i}^{y}S_{j}^{y}+\Delta S_{i}^{z}S_{j}^{z} ,\quad
   \Delta>0 , \label{XXZ}
 \end{equation}
 and $J_1$ denotes the intra-trimer coupling, and $J_2$ and $J_3$ the
 inter-trimer couplings.  All the couplings are supposed to be
 antiferromagnetic ($J_1$,$J_2$,$J_3>0$).  Okamoto and Ichikawa\cite{Okamoto1}
 discussed the case of $\Delta_{1}=\Delta_{2}=\Delta_{3}$.  We
 use the generalized version to investigate the relation between the
 anisotropy parameter and the inversion phenomenon.  Hereafter we
 consider a case, where $J_{1}\gg J_{2},J_{3}$.
 \section{Phase diagram}
\begin{figure}
  \begin{center}
    \scalebox{0.3}[0.3]{\includegraphics{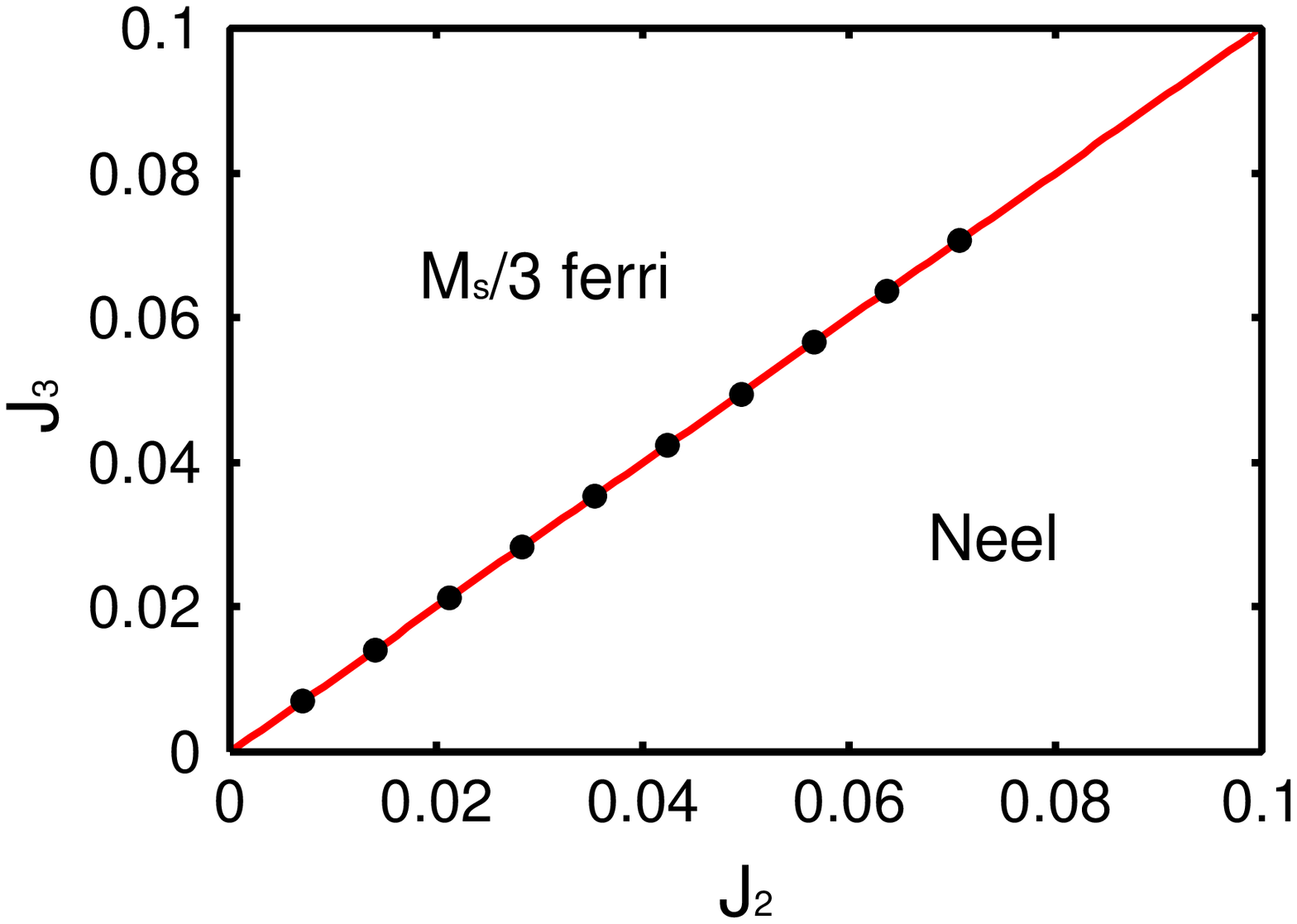}}
    \caption{A phase diagram for $J_1=1,
    \Delta_1=1, \Delta_2=\Delta_3=2.5$.  Solid line denotes the DPT
    result, and dots denote the numerical result.}
    \label{Isinglike1}
    \scalebox{0.3}[0.3]{\includegraphics{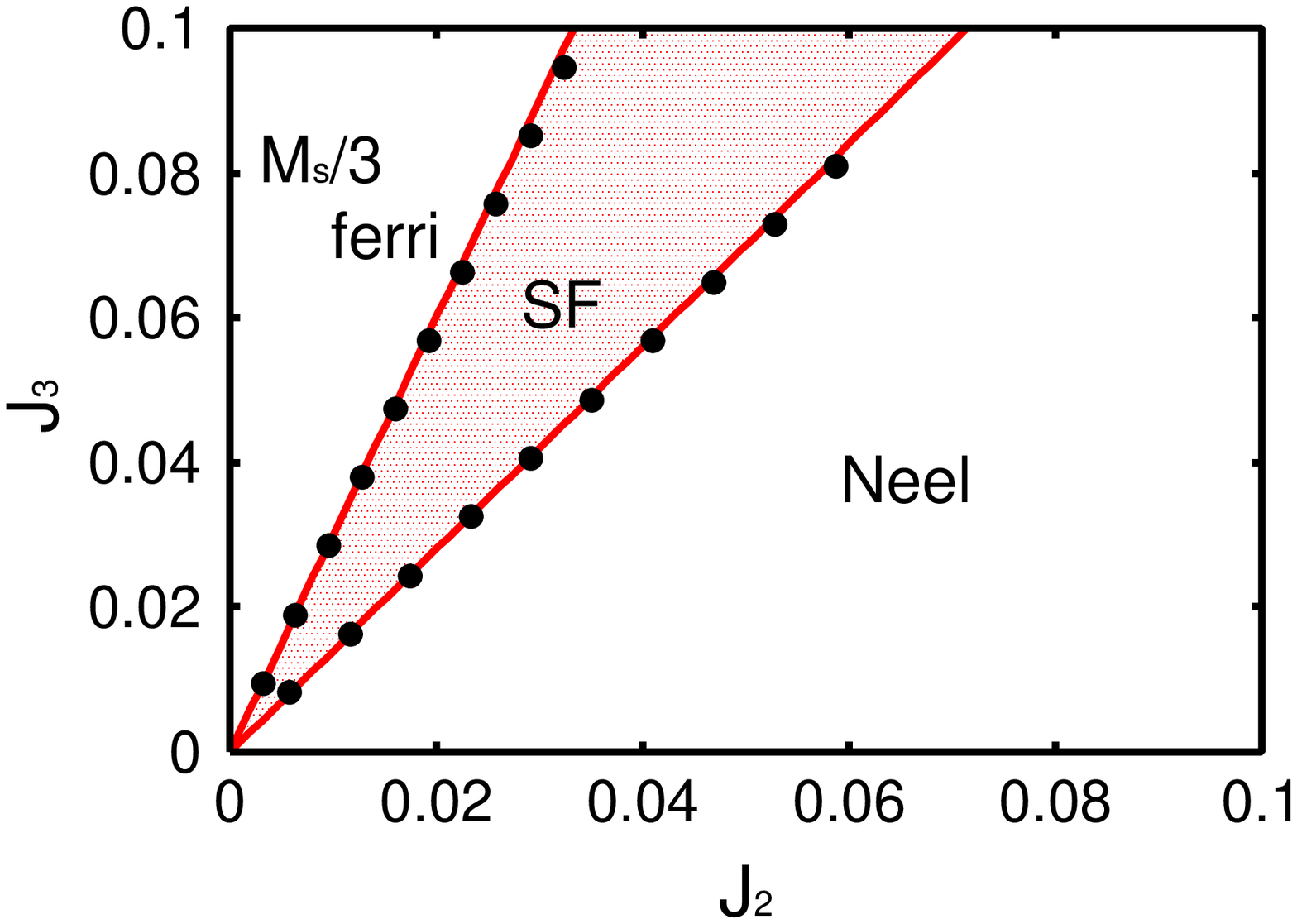}}
    \caption{A phase diagram for $J_1=1,
    \Delta_1=1, \Delta_2=2.5, \Delta_3=1.5$.  Solid lines denote the DPT
    result, and dots denote the numerical result.}
    \label{Isinglike2}
    \scalebox{0.3}[0.3]{\includegraphics{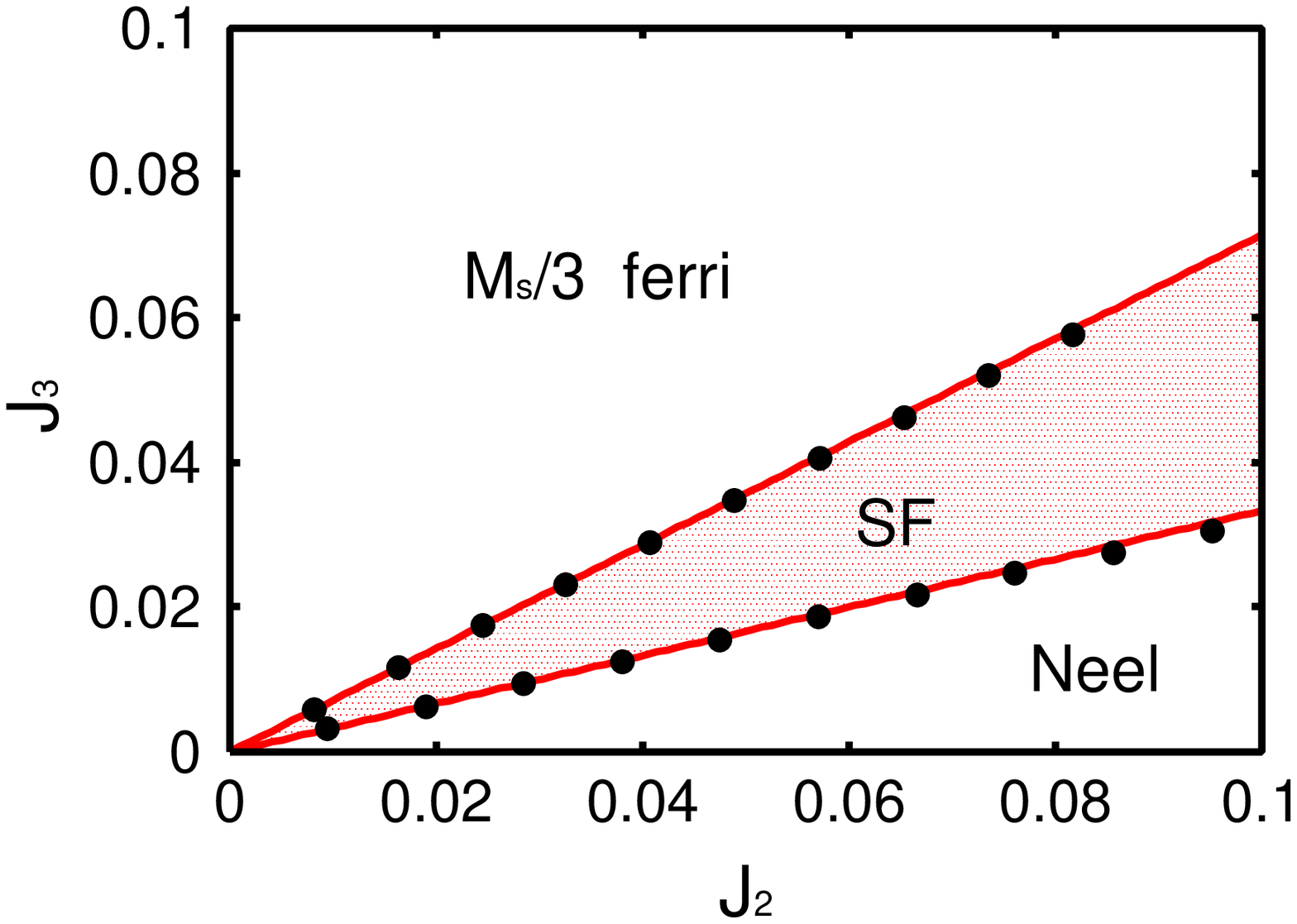}}
    \caption{A phase diagram for $J_1=1,
    \Delta_1=1, \Delta_2=1.5, \Delta_3=2.5$.  Solid lines denote the DPT
    result, and dots denote the numerical result.}
    \label{Isinglike3}
  \end{center}
 \end{figure}

\begin{figure}
  \begin{center}
    \scalebox{0.3}[0.3]{\includegraphics{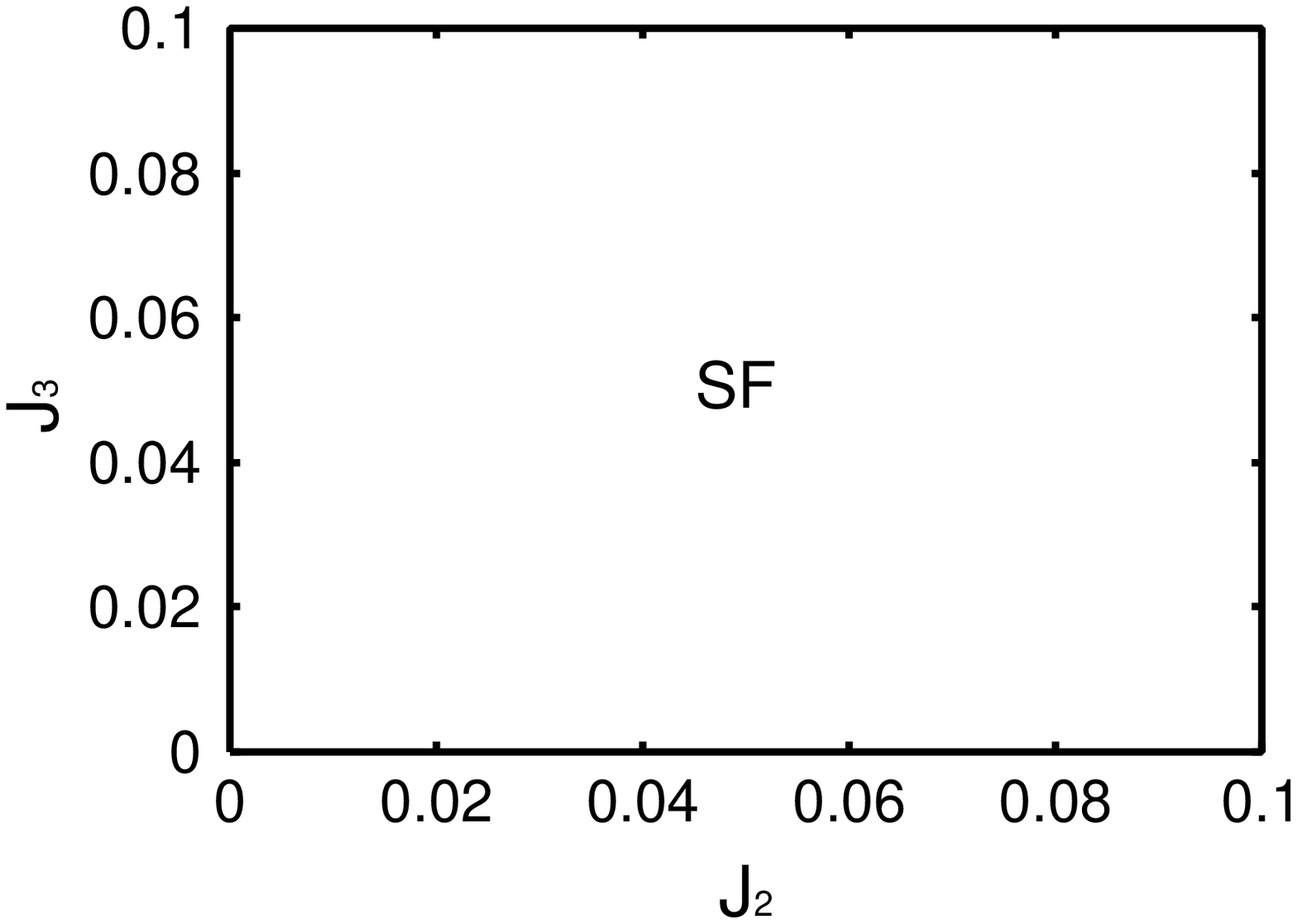}}
    \caption{A phase diagram for $J_1=1, \Delta_1=1, \Delta_2=\Delta_3=0.5$.}
    \label{XYlike1}
    \scalebox{0.3}[0.3]{\includegraphics{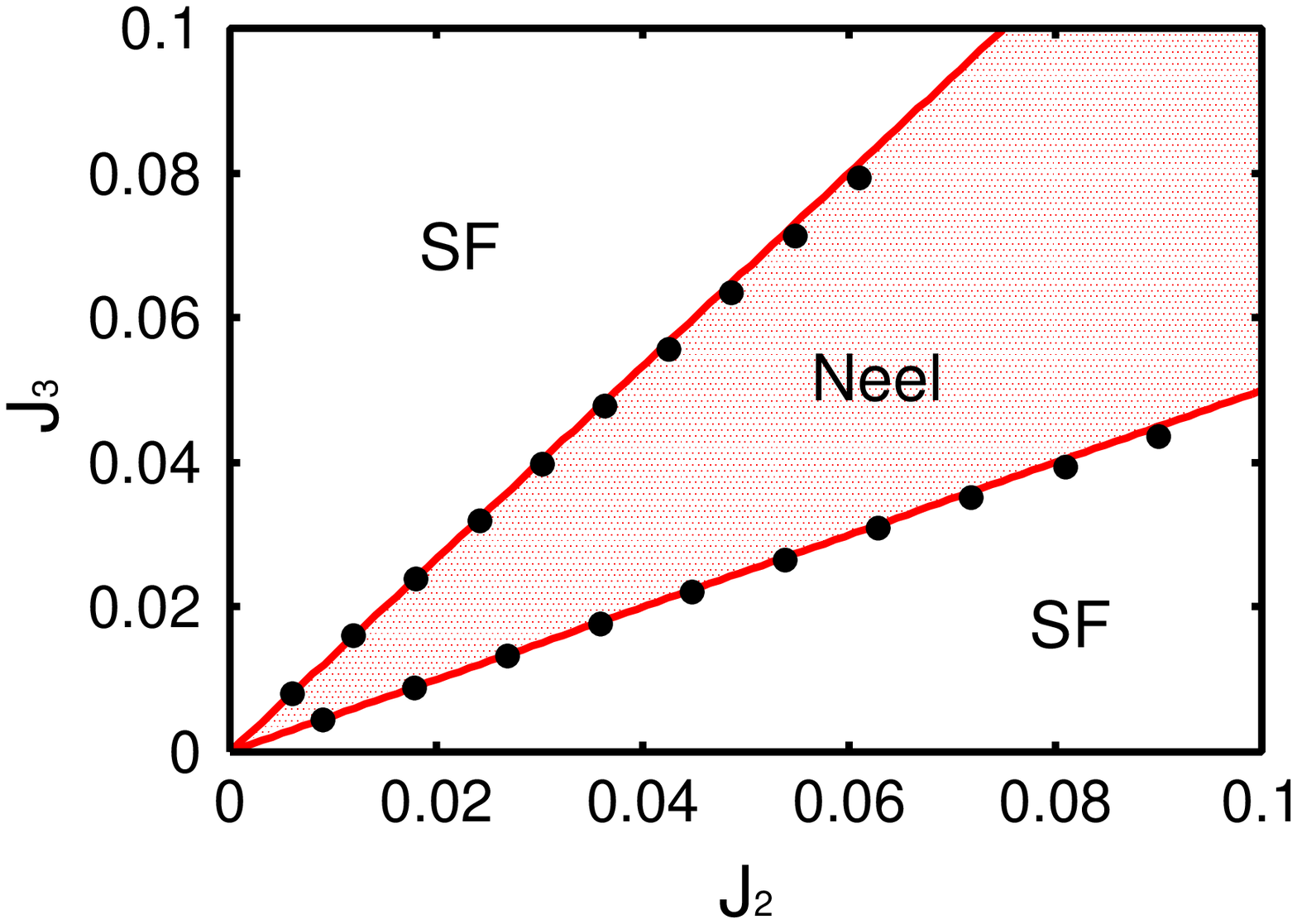}}
    \caption{A phase diagram for $J_1=1, \Delta_1=1, \Delta_2=0.6,
    \Delta_3=0.2$.  Solid lines denote the DPT result, and dots denote
    the numerical result.}
    \label{XYlike2}
    \scalebox{0.3}[0.3]{\includegraphics{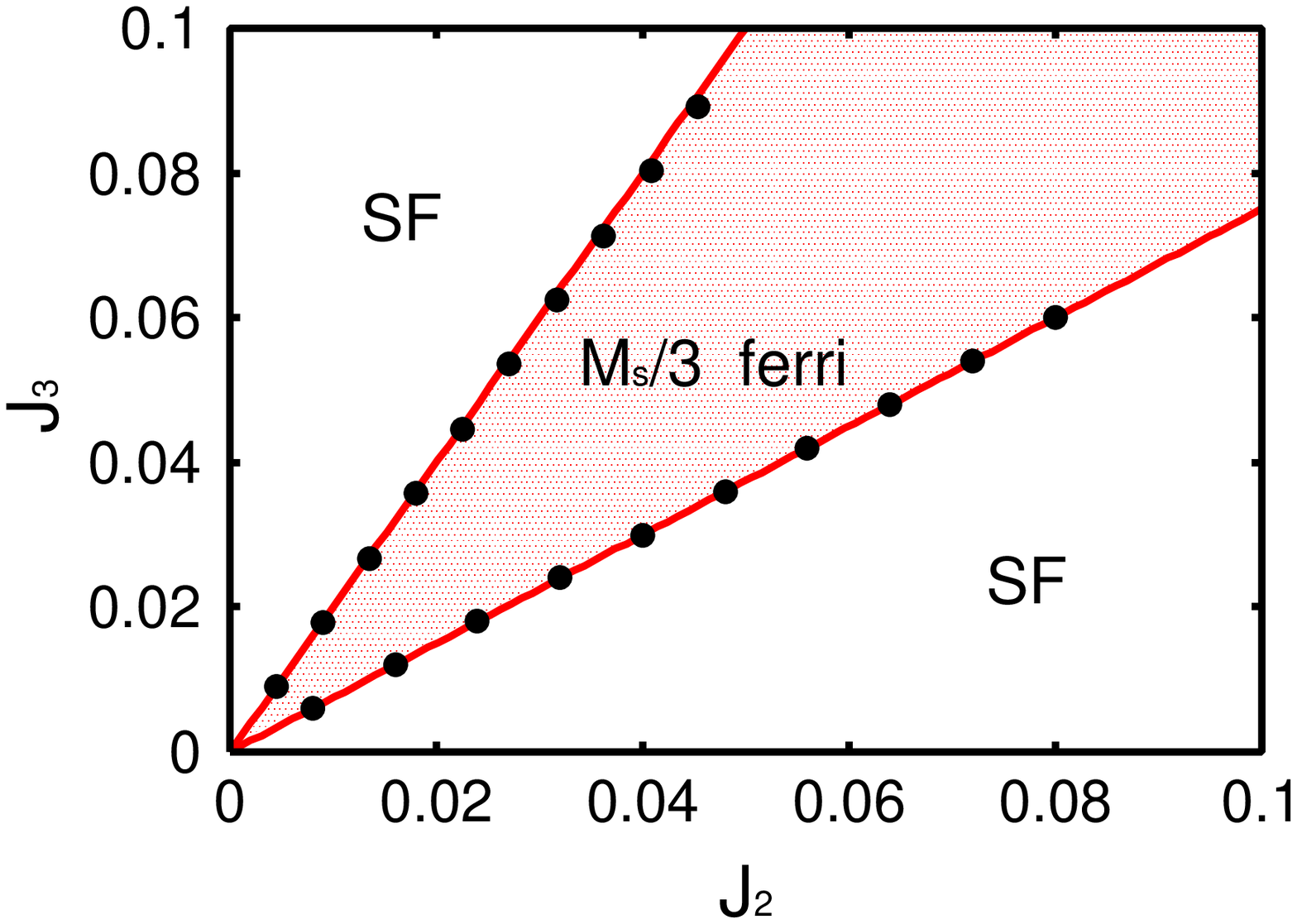}}
    \caption{A phase diagram for $J_1=1, \Delta_1=1, \Delta_2=0.2,
    \Delta_3=0.6$.  Solid lines denote the DPT result, and dots denote
    the numerical result.}
    \label{XYlike3}
  \end{center}
 \end{figure}

 Let us discuss the ground state of the Hamiltonian Eq.(\ref{H}) using
 the degenerate perturbation theory\cite{DPT1,DPT2,DD2} (DPT) up to the
 first order.  First, we
 consider a $J_{2}=J_{3}=0$ case which is merely a three-spin
 problem.  The ground states of the $j$-th trimer are
 \begin{eqnarray}
   |\Uparrow_{j}\rangle &\equiv
   \frac{1}{\sqrt{2+\lambda^2}}\left(|\uparrow\uparrow\downarrow\rangle-\lambda|\uparrow\downarrow\uparrow\rangle+|\downarrow\uparrow\uparrow\rangle\right)
    (S_{\rm tot}^z=1/2) ,\\
   |\Downarrow_{j}\rangle &\equiv
   \frac{1}{\sqrt{2+\lambda^2}}\left(|\uparrow\downarrow\downarrow\rangle-\lambda|\downarrow\uparrow\downarrow\rangle+|\downarrow\downarrow\uparrow\rangle\right)
    (S_{\rm tot}^z=-1/2),
 \end{eqnarray}
 where $|\uparrow \uparrow \downarrow \rangle$ denotes $|\uparrow
 \rangle_{3j-1} \otimes |\uparrow \rangle_{3j} \otimes |\downarrow
 \rangle_{3j+1}$, and $\lambda$ is a constant
 \begin{equation}
  \lambda=\frac{\Delta_{1}+\sqrt{\Delta_{1}^{2}+8}}{2} >\sqrt{2} \quad (\Delta_{1}>0).
 \end{equation}
 Expectation values of the $z$-component of the spin are 
 \begin{eqnarray}
  \langle\Uparrow_j|S_{3j\pm1}^z|\Uparrow_j\rangle=-\langle\Downarrow_j|S_{3j\pm1}^z|\Downarrow_j\rangle=\frac{\lambda^2}{2(\lambda^2+2)}>0 , \label{expectS1}\\
  \langle\Uparrow_j|S_{3j}^z|\Uparrow_j\rangle=-\langle\Downarrow_j|S_{3j}^z|\Downarrow_j\rangle=-\frac{\lambda^2-2}{2(\lambda^2+2)}<0 \label{expectS2}.
 \end{eqnarray} 
 An eigenstate of $N$-trimer system is a tensor product of an
 eigenstate of the $j$-th trimer Hamiltonian.  Since the ground states
 of a trimer are doubly degenerate, the
 ground states $|\Psi \rangle$ of $N$-trimer system are $2^{N}$-fold degenerate
 and they are
 \begin{equation}
  |\Psi \rangle = \bigotimes_{j}|\psi_{j}\rangle ,\quad |\psi_{j}\rangle=
   |\Uparrow_j\rangle \  \mbox{or} \  |\Downarrow_j\rangle .
 \end{equation}
 Next, we consider the $J_{1} \gg J_{2},J_{3}$ case, where $J_2$ and $J_3$
 terms can be treated as perturbations.  As far as we consider only the
 first-order corrections of the perturbations, we regard $|\Psi\rangle$ as
 a basis of a whole Hilbert space.  It is convenient to introduce the
 pseudo-spin operator $\vT_j$ with $T_j=1/2$, by which
 $|\Uparrow_j\rangle$ and $|\Downarrow_j\rangle$ are expressed as
 $T_j^z=1/2$ and $T_j^z=-1/2$ states, respectively.  The original spin
 operators can be represented by the pseudo-spin operators:
 \begin{eqnarray}
  & S_{3j\pm1}^x \equiv -\frac{2\lambda}{2+\lambda^2}T_j^x ,\quad S_{3j\pm1}^y \equiv -\frac{2\lambda}{2+\lambda^2}T_j^y ,\\
  & S_{3j\pm1}^z \equiv \frac{\lambda^2}{2+\lambda^2}T_j^z ,\\
  & S_{3j}^x \equiv \frac{2}{2+\lambda^2}T_j^x ,\quad S_{3j}^y \equiv \frac{2}{2+\lambda^2}T_j^y ,\\
  & S_{3j}^z \equiv \frac{2-\lambda^2}{2+\lambda^2}T_j^z,
 \end{eqnarray}
 within the reduced basis.  By straightforward calculations,
 the first-order correction with respect to $J_{2}$ and $J_{3}$ can be
 obtained as
 \begin{equation}
  {\cal H}_{\rm eff} = \sum_{j}[J_{\rm
   eff}^{\perp}(T_{j}^{x}T_{j+1}^{x}+T_{j}^{y}T_{j+1}^{y})+J_{\rm
   eff}^{z}T_{j}^{z}T_{j+1}^{z}] \label{Heff},
 \end{equation}
 where
 \begin{eqnarray}
   J_{\rm eff}^{\perp}&=&\frac{4\lambda(\lambda
   J_{2}-2J_{3})}{(2+\lambda^2)^2} , \label{effJperp} \\
   J_{\rm
    eff}^{z}&=&\frac{\lambda^2\{\lambda^2\Delta_{2}J_{2}-2(\lambda^2-2)\Delta_{3}J_{3})\}}{(2+\lambda^2)^2} . \label{effJz}
 \end{eqnarray}
 In Eq.(\ref{Heff}), constant terms are omitted.  This effective model is
 nothing but the one-dimensional spin-$1/2$ $XXZ$ model.  The $XXZ$
 model has three kinds of the ground states depending on the anisotropy: the
 ferromagnetic state, the SF state, and the N\'{e}el state.  The
 ferromagnetic state in the $\vT$-picture corresponds to the
 $M_{s}/3$ ferrimagnetic state in the $\vS$-picture, where
 $M_{s}$ is the saturation magnetization. Thus, the
 ground state of the original system is obtained from $J_{\rm
 eff}^{\perp}$ and $J_{\rm eff}^{z}$ as follows.
 \begin{equation*}
  \begin{array}{c c l}
   J_{\rm eff}^{z}<0 \ \mbox{and}\ |J_{\rm eff}^{z}|>|J_{\rm eff}^{\perp}| &
    \Longrightarrow & \mbox{ferrimagnetic state} \\
   |J_{\rm eff}^{z}|<|J_{\rm eff}^{\perp}| &
    \Longrightarrow & \mbox{spin-fluid state} \\
   J_{\rm eff}^{z}>|J_{\rm eff}^{\perp}| &
    \Longrightarrow & \mbox{N\'{e}el state} \\
  \end{array}
 \end{equation*}
 From the above inequalities, the
 phase diagram of the ground state with respect to three variables,
 $\Delta_{1}$, $\Delta_{2}$, and $\Delta_{3}$, can be obtained.  The examples of
 the phase diagram are shown in Figs.\ref{Isinglike1}-\ref{XYlike3}.
 Figures \ref{Isinglike1}-\ref{Isinglike3} are examples of the case in which the
 anisotropy parameters, $\Delta_{1}$, $\Delta_{2}$, and $\Delta_{3}$, are
 Ising-like.  Figures \ref{XYlike1}-\ref{XYlike3} are examples of the case
 of the $XY$-like anisotropy parameters.  Although the phase diagrams
 in a $\Delta_1 \ne 1$ case are not shown in this paper, the inversion
 phase also appears in that case.  From
 Figs.\ref{Isinglike1}-\ref{XYlike3}, it is clear that the inversion
 phases generally appear. Only in the cases of
 $\Delta_{1}=1,\Delta_{2}=\Delta_{3}$, however, the inversion phases do
 not appear. 

 As a numerical check, we have also performed the numerical
 diagonalization using the Lancz\"{o}s method and the level
 spectroscopy analysis.\cite{LS1,LS2,LS3,LS4}   We found that
 the numerical results are in good agreement with the results of the
 perturbation theory as shown in Figs.\ref{Isinglike1}-\ref{XYlike3}.
 \section{Discussion}
\begin{figure}
  \begin{center}
   \scalebox{0.2}[0.2]{\includegraphics{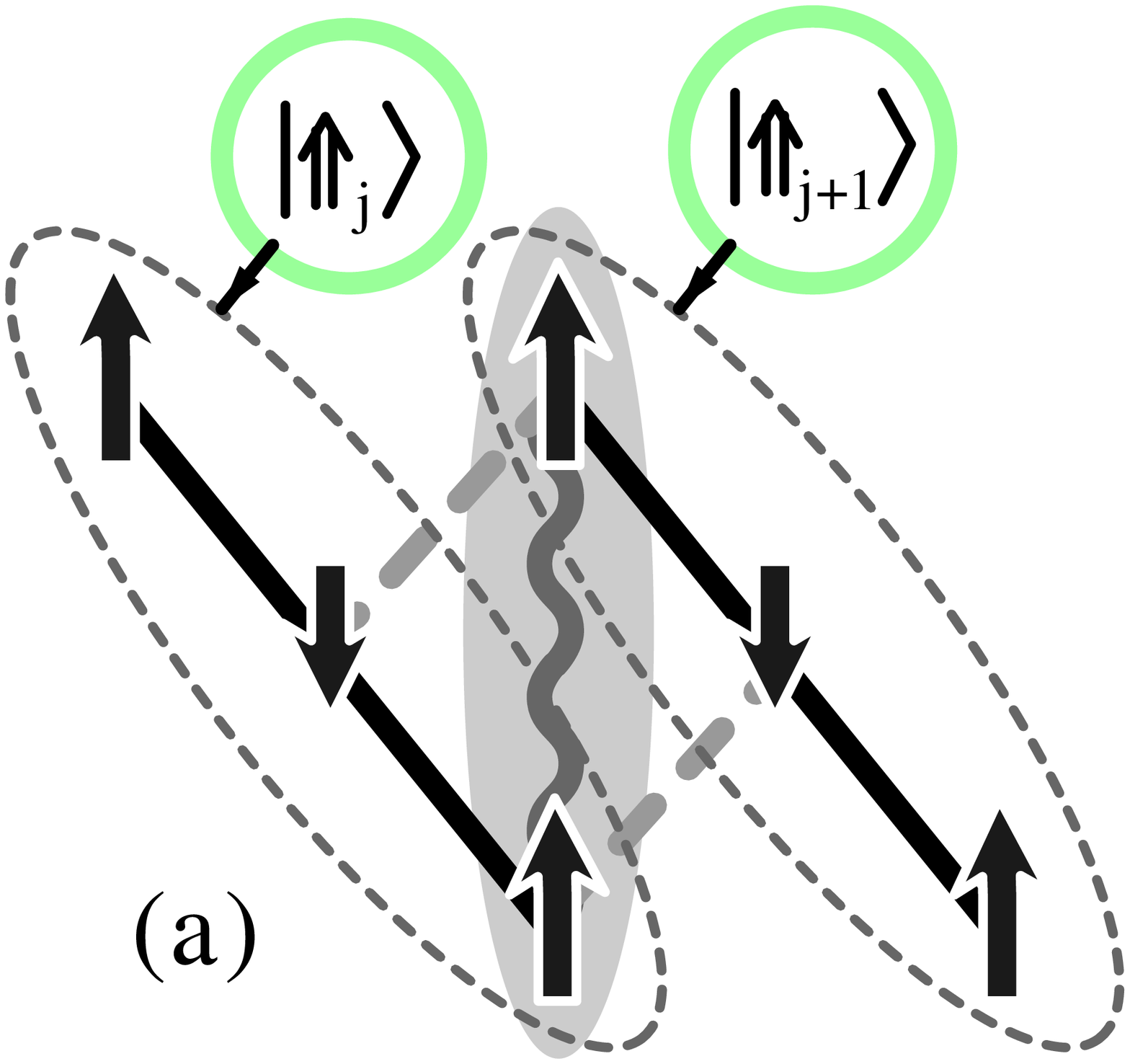}} 
   \scalebox{0.2}[0.2]{\includegraphics{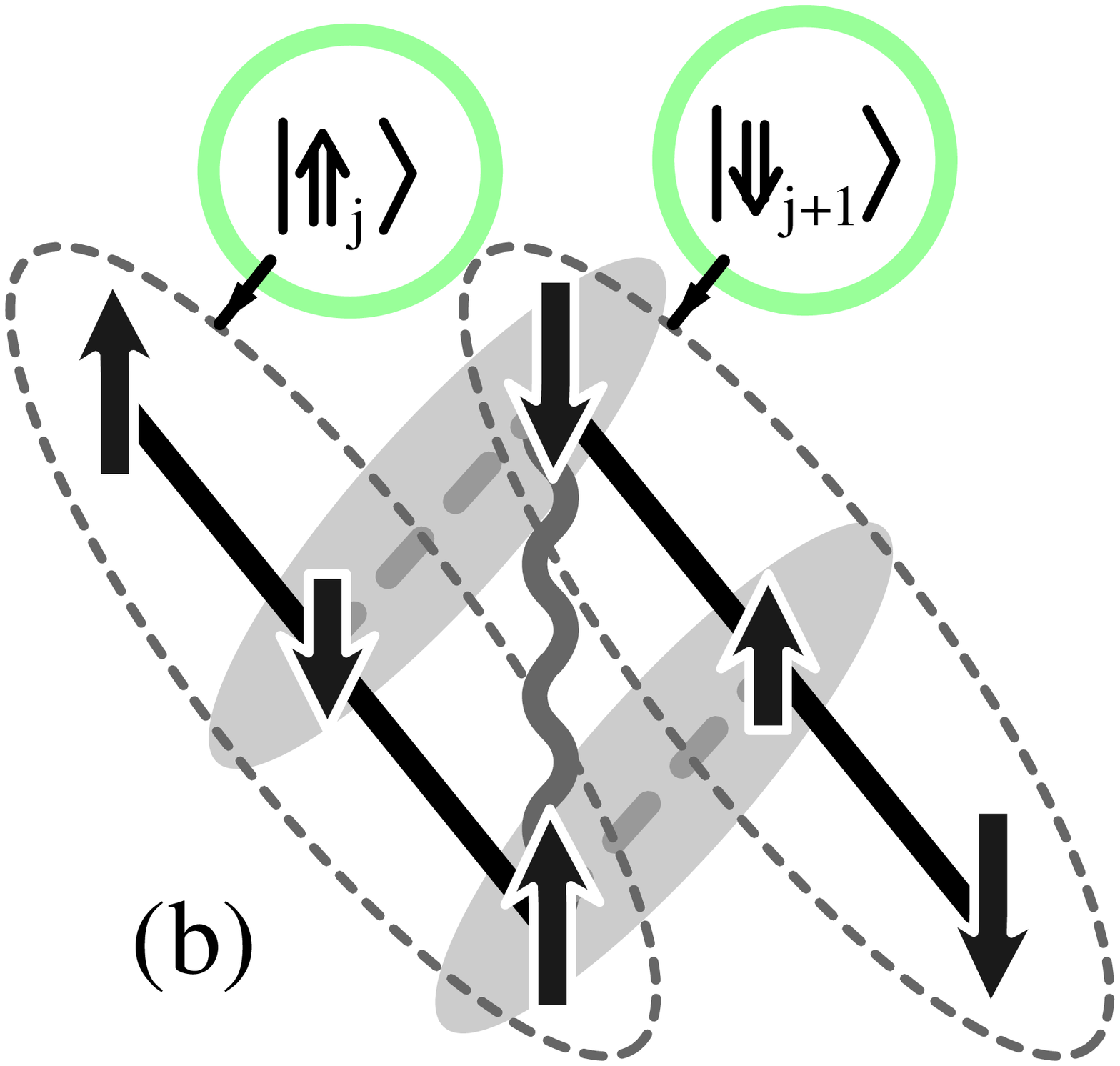}}
   \caption{Schematic diagrams for the cases where two trimers are
   neighboring each other with (a) $|\Uparrow_j\rangle$ and
   $|\Uparrow_{j+1}\rangle$, and (b) $|\Uparrow_j\rangle$ and
   $|\Downarrow_{j+1}\rangle$.  The shadow denotes a frustrated
   pair of spins.}
   \label{alignment}
  \end{center}
 \end{figure}

\begin{figure}
  \begin{center}
   \scalebox{0.3}[0.3]{\includegraphics{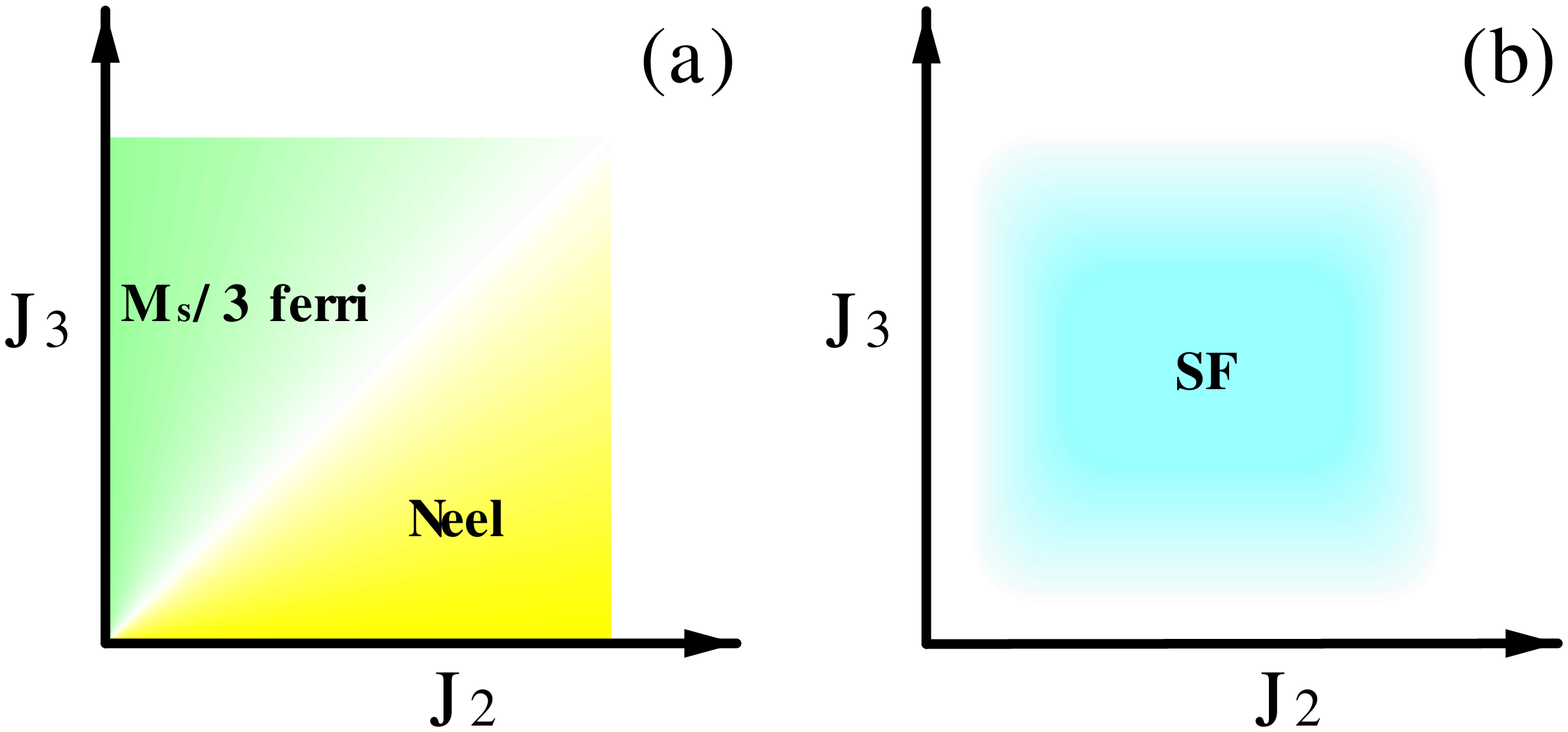}}
   \caption{Schematic phase diagrams with the qualitative argument
   for (a)$\Delta_2,\Delta_3>1$ (Ising-like) case, and (b)$\Delta_2,\Delta_3<1$ ($XY$-like) case.}
   \label{PD}
  \end{center}
 \end{figure}

 Let us discuss the ground-state phase diagram of the DD chain qualitatively.
 First, we consider a case where $\Delta_2,\Delta_3>1$.  We note that the
 following discussions are qualitatively independent of
 $\Delta_1$.  We discuss a correlation of neighboring trimers.
 Figure \ref{alignment}(a) shows a case where the two trimers are
 neighboring each other with $|\Uparrow_j\rangle$ and $|\Uparrow_{j+1}\rangle$.
 Figure \ref{alignment}(b) shows a case where they are
 neighboring with $|\Uparrow_j\rangle$ and
 $|\Downarrow_{j+1}\rangle$.  The couplings $J_2$ is
 frustrated for Fig.\ref{alignment}(a) since all the interactions are
 antiferromagnetic. Thus, we can consider that the situation of 
 Fig.\ref{alignment}(a) is stable for larger $J_3$, while it is unstable
 for larger $J_2$.  On the other hand, the coupling $J_3$ is frustrated for
 Fig.\ref{alignment}(b).  Thus, we can
 also consider that the situation of Fig.\ref{alignment}(b) is stable for
 larger $J_2$, and unstable for larger $J_3$.  Of course, we did not
 mention that the phase boundary is $J_2=J_3$ because its slope will
 depend on the values of anisotropies, as is expected from
 Eqs.(\ref{expectS1})-(\ref{expectS2}).  We obtain the qualitative
 sketch of the phase diagram
 shown in Fig.\ref{PD}(a); the area of $J_2<J_3$ is the
 ferrimagnetic phase and that of $J_2>J_3$ is the N\'{e}el phase.

 Next, we discuss a case of $\Delta_2,\Delta_3<1$.  The ordering
 effect of the couplings  $J_2$ and $J_3$ are weak,
 because the $\Delta_2$ and $\Delta_3$ are $XY$-like.  Therefore,
 a whole area of the $J_2$-$J_3$ phase diagram where $J_2, J_3 \ll
 J_1$ will be the SF state (see Fig.\ref{PD}(b)).
 Comparing Fig.\ref{PD} with Figs.\ref{Isinglike1}-\ref{XYlike3}, we can see
 that the above discussions hit the point. 
   
 Let us proceed to the inversion phase.  For
 Figs.\ref{Isinglike2}-\ref{Isinglike3} and
 Figs.\ref{XYlike2}-\ref{XYlike3}, the inversion phase appears in 
 an area $J_1 \gg J_2 \sim J_3$
 where the trimer nature is still conspicuous,
 and $J_2$ and $J_3$ strongly compete with each other.
 We present an
 intuitive explanation in this competitive area.  The interaction
 along the $z$-direction is stronger for the Ising-like case,
 by which the system is well-ordered to the N\'{e}el state in usual cases.
 However, the ordered state undergoes more energy loss in this area
 due to the competition between $J_2$ and $J_3$.
 Thus, the spins are going to lie in the $XY$ plane to avoid this 
 energy loss, which brings about the disordered SF phase.
 A similar explanation holds for the $XY$-like case.
 The spins are going to lie in the $XY$-plane in usual cases,
 which corresponds to the SF state.
 In our case, since the energy loss due to the competition between $J_2$ and $J_3$ is serious,
 the spins are going to be pointed along the $z$-direction.
 Then, the ordered state (N\'{e}el or ferrimagnetic state) is realized.
 Another possible mechanism to avoid the energy loss due to the competing interactions 
 is to form a singlet dimer pair.
 This mechanism is well known for the $S=1/2$ chain with the
 next-nearest-neighbor interactions.\cite{TH,ON}  However, the
 formation of the singlet dimer pair is irreconcilable with the trimer nature. 
 Therefore, the dimer phase does not appear for $J_1 \gg J_2,J_3$,
 where the trimer nature is conspicuous.
 In fact, we have found that the dimer phase appears in the regime $J_1 \sim J_2 \sim J_3$.\cite{OTY}
 Thus, we have intuitively explained why the inversion phenomenon appears in this area.  
   
 Thereby, it seems that the origin of the inversion phenomenon is frustration,
 the trimer nature, and the $XXZ$ anisotropy.  
 In addition, the above discussion can also
 predict the location of the inversion phase.  The inversion phase in
 Fig.\ref{Isinglike2} is located on the upper left side, compared to that in
 Fig.\ref{Isinglike3}.  This is because of the magnitude relation between
 $\Delta_2$ and $\Delta_3$ as explained in the following.
 In the above discussion, only the
 couplings $J_2$ and $J_3$ are considered as the ordering effect.
 However, not only $J_2$ and $J_3$, but also $\Delta_2$
 and $\Delta_3$ govern the orders.  When $\Delta_2>\Delta_3$, the
 N\'{e}el state area becomes wider in
 Fig.\ref{Isinglike2}, and the inversion phase is located on the upper left side.
 On the other hand, the ferrimagnetic state area becomes wider in
 Fig.\ref{Isinglike3} because $\Delta_2<\Delta_3$; this helps for the role
 of $J_3$.  Comparing Figs.\ref{XYlike2}-\ref{XYlike3}, we can also
 develop a similar discussion.  The inversion phase is the N\'{e}el
 phase in Fig.\ref{XYlike2}, and it is the ferrimagnetic phase in
 Fig.\ref{XYlike3}.  When $J_2$ and $\Delta_2$ ($J_3$ and $\Delta_3$) are relatively
 strong in comparison with $J_3$ and $\Delta_3$ ($J_2$ and $\Delta_2$),
 the more stable state is Fig.\ref{PD}(b) (Fig.\ref{PD}(a)).  Thus, the N\'{e}el ordering
 effect is stronger for Fig.\ref{XYlike2}, and the ferrimagnetic phase appears for
 Fig.\ref{XYlike3}.  We can interpret that the N\'{e}el phase
 (the ferrimagnetic phase) is selected as the inversion phase in
 Fig.\ref{XYlike2} (Fig.\ref{XYlike3}).

 The phase diagrams of the cases of Fig.\ref{Isinglike1} and Fig.\ref{XYlike1}
 are remarkable.  The inversion phase does not appear in these phase
 diagrams.  The result of the analytical calculation indicates that the
 inversion phases are always absent when $\Delta_1=1$ and
 $\Delta_2=\Delta_3$.  The physical reason of the absence of the
 inversion phenomenon has not been understood yet.

 Finally, we consider the origin of the inversion phenomenon in view of
 the effective model, Eqs.(\ref{Heff})-(\ref{effJz}).  The effective
 coupling constants $J_{\rm eff}^{\perp}$ and $J_{\rm eff}^{z}$ are
 constructed of the subtraction of the terms proportional to $J_2$
 and $J_3$.  Of course, the subtraction comes from the competition
 between couplings $J_2$ and $J_3$.  It can be manifestly
 expected that this subtraction causes the inversion between the
 anisotropy of the original model and that of the effective model.
 For instance, we discuss the case of $\Delta_1=1$, where $J_{\rm
 eff}^z/J_{\rm eff}^{\perp}=(\Delta_2J_2-\Delta_3J_3)/(J_2-J_3)$.
 When $J_2 \gg J_3$, we see $J_{\rm eff}^z/J_{\rm eff}^{\perp} \sim \Delta_2$ and $J_{\rm
 eff}^{z}>0$.  Meanwhile, when $J_2 \ll J_3$, it becomes $J_{\rm
 eff}^z/J_{\rm eff}^{\perp} \sim \Delta_3$ and $J_{\rm
 eff}^{z}<0$.  Thus, we obtain the schematic phase diagram
 Fig.\ref{PD}.  This agrees with the result of the previous
 qualitative discussion.  On the other hand, when $J_2 \sim
 J_3$, $J_{\rm eff}^z/J_{\rm eff}^{\perp}$ can be Ising-like
 ($XY$-like) in spite of the $XY$-like (Ising-like) $\Delta_2$ and
 $\Delta_3$.  Here the subtraction plays a crucial role.

 \section{Conclusion}
 In this paper, we investigate the phase diagram of the ground state of
 the DD chain with the $XXZ$ type interaction using the degenerate
 perturbation theory as well as the level spectroscopy analysis of the
 numerical diagonalization data.  The obtained phase diagrams indicate
 that the inversion phase is stable to various sets of anisotropies.  Our discussion
 suggests that frustration, the trimer nature and the $XXZ$ type
 interaction are necessary for the inversion phenomenon.  In fact, it
 has been found in the $S=1/2$ trimerized $XXZ$ chain
 with the next-nearest-neighbor interactions,\cite{Okamoto2} and also
 in the $S=1/2$ frustrated three-leg ladder with the $XXZ$
 anisotropy. \cite{Okamoto-3leg}  Therefore, we conclude that the
 inversion phase may appear in the
 phase diagram of the ground state for models with these three key
 words.  Unfortunately, real materials corresponding to
 the above models have not been found yet.  However, our prediction of
 this interesting phenomenon may become a motivation of searching or
 synthesizing the corresponding materials.

 \section{Acknowledgments}
 We would like to express our appreciation to Prof. Takashi Tonegawa
 and Prof. Masaki Oshikawa for stimulating discussions.  For the
 numerical diagonalization, we used the package program TITPACK
 ver.2 developed by Prof. Hidetoshi Nishimori, to whom we are grateful.
 The present work has been supported in part by a Grant-in-Aid for
 Scientific Research (C) (No.14540329) from the Ministry of Education,
 Culture, Sports, Science and Technology.


\section*{References}
  \begin{thebibliography}{99}
   \bibitem{Okamoto1} K. Okamoto and Y. Ichikawa: J. Phys. Chem. Solids
   {\bf 63} (2002) 1575.	  
   \bibitem{Okamoto2} K. Okamoto: Prog. Theor. Phys, Suppl. No. 145
   (2002) 208.
   \bibitem{DD1} K. Okamoto, T. Tonegawa, Y. Takahashi and M. Kaburagi:
   J. Phys. : Cond. Matt. {\bf 11} (1999) 10485.
   \bibitem{DD2} A. Honecker and A. L\"{a}uchli: Phys. Rev. B {\bf 63}
   (2001) 174407.
   \bibitem{DD3} K. Okamoto, T. Tonegawa and M. Kaburagi: J. Phys. :
   Cond. Matt. {\bf 15} (2003) 5979, and references therein.
   \bibitem{DPT1} K. Totsuka: Phys. Rev. B {\bf 57} (1998) 3435.
   \bibitem{DPT2} F. Mila: Eur. Phys. J. B {\bf 6} (1998) 201.
   \bibitem{LS1} K. Nomura and A. Kitazawa: in {\it Proc. French-Japanese
   Symp. on Quantum Properties of Low-Dimensional
   Antiferromagnets} ed Y. Ajiro and J-P. Bouchcer (Kyushuu
   University Press, 2002); cond-mat/0201072.
   \bibitem{LS2} K. Okamoto: Prog. Theor. Phys, Suppl. No. 145 (2002)
   113; cond-mat/0201013.
   \bibitem{LS3} K. Nomura and K. Okamoto: J. Phys. Soc. Jpn. {\bf 62}
   (1993) 1123.
   \bibitem{LS4} K. Nomura and K. Okamoto: J. Phys. A: Math. Gen. {\bf
   27} (1994) 5773.
   \bibitem{TH} T. Tonegawa and I. Harada: J. Phys. Soc. Jpn. 56 (1987)
   2153.
   \bibitem{ON}K. Okamoto and K. Nomura: Phys. Lett. A 169 (1992) 433.
   \bibitem{OTY} K. Okamoto, A. Tokuno and Y. Ichikawa: in preparation.
   \bibitem{Okamoto-3leg} K. Okamoto and T. Sakai: in preparation.
  \end{thebibliography}
 \end{document}